**Dirac's fine-structure formula for an "abnormally high" nuclear charge**
*The solution of an old riddle*

A. LOINGER

Dipartimento di Fisica dell'Università di Milano

Via Celoria, 16 –  20133 Milano, Italy

**Summary.** – In Dirac's fine-structure formula for the hydrogenlike atoms a critical role is played by the square root of the following expression:  the unity minus the square of the product of the atomic number by the fine-structure constant (which is approximately equal to 1/137). Indeed, for a fictitious, ideal nucleus for which the above product is greater than or equal to one, the mentioned square root becomes imaginary, or zero. I prove in this Note that the origin of such theoretical "breaking down" is quite simple and is inherent in the special theory of relativity of *classical* physics.

**1. – The problem**

As is well known, Dirac's fine-structure formula for the energy of an electron in the Coulomb field generated by a fixed nuclear charge *Ze* is as follows:

$$(1.1) \qquad \frac{E}{m_0 c^2} = \left[ 1 + \frac{Z^2 \alpha^2}{\left[ n_r + (\kappa^2 - Z^2 \alpha^2)^{1/2} \right]^2} \right]^{-1/2},$$

where: $E = W + m_0 c^2$ is the relativistic total energy; $m_0$ is the rest-mass of the electron; $\alpha := e^2/(\hbar c)$ is the fine-structure constant; $n_r = 0,1,2,3,...$ is the radial quantum number; $\kappa = \pm 1, \pm 2, \pm 3,...$ is the auxiliary quantum number (according to a Weylian terminology); when $n_r = 0$, the quantum $\kappa$ takes only the positive values [1]. (Remark that Dirac writes $j$ in lieu of $\kappa$, but Dirac's $j$ does not coincide with the inner quantum number, which is usually denoted with $j$. For the customary $j$ we have $j = |\kappa| - 1/2$ ). Weyl [1] emphasizes that the number of components corresponding to the fine-structure formula (1.1) is greater than in Sommerfeld's theory [2]: indeed, in addition to the transitions $\kappa \to \kappa - 1$, and $\kappa \to \kappa + 1$, we may also have $\kappa \to -\kappa$, an addition which is in agreement with the experiments [3].

For a fictitious nuclear charge $Z_* e$ and auxiliary quantum numbers $\kappa_*$ such that $\kappa_*^2 - Z_*^2 \alpha^2 \leq 0$, eq. (1.1) loses obviously any *physical* meaning. What's the significance (if any) of this "failure"? (Remark that in Schrödinger nonrelativistic theory there is no limitation of this kind).

In the past half-century many physicists have tried to find a reasonable answer, but with a scarce success. One of them "resolved" the difficulty by requiring all eigenfunctions to approach the origin of the coordinate system with the *same* constant phase, *arbitrarily chosen*. In this way he created a series of eigenvalues free from the above limitation. However, the overwhelming majority of the authors followed the line of thought proposed, e.g., by Bjorken and Drell [4], who remarked that "for $Z\alpha \geq 1$, $(1 - Z^2\alpha^2)^{1/2}$ is imaginary [or zero] and the solutions [of Dirac equation] exhibit an oscillatory behavior reminiscent of that found in the Klein paradox." [5]. This observation suggests that the explanation of the riddle ought to be sought in the positron theory of quantum electrodynamics. Nevertheless, no convincing answer has been found from this viewpoint – and *pour cause*, as we shall see.

I prove in the sequel that the problem can be solved in a plain and rational way, the root of the enigma being inherent in the special theory of relativity.

## 2. – **The solution**

The energies $E$ of the discrete levels of a relativistic *one*-particle system are restricted to the domain between $-m_0c^2$ and $+m_0c^2$, whereas for the energies of the scattering states we have $E \geq +m_0c^2$ or $E \leq -m_0c^2$. Of course, from the *physical* standpoint we select for the bound states the energies $E$ between $0$ and $+m_0c^2$, and for the scattering states the energies $E$ such that $E \geq +m_0c^2$, see e.g. Dirac [1].

In 1932 Pauli applied the WKB-method to Dirac equation [6], and showed that: *i*) the diffraction effects of the electron waves and the spin actions have the *same* order of magnitude, thus confirming an important thesis by Bohr; *ii*) the rays of the "geometrical optics", which follows from the "wave optics" corresponding to Dirac equation, coincide with the trajectories of the classical relativistic dynamics of point particles *without* spin. (This theorem can be further corroborated by means of the mathematical theory of the characteristics).

Pauli's results assign a precise significance to the fine-structure formula for the hydrogenlike atoms discovered by Sommerfeld in 1916 [2]: by means of Sommerfeld-Wilson conditions, this Author selected a discrete subset of the *classical* relativistic orbits $(0 < E < m_0c^2)$ for a point electron *without* spin. We can say that Sommerfeld's formula gives just the right Bohr-Sommerfeld approximation to Dirac's fine-structure formula, which concerns an electron *with* spin (a spin generated by the *Zitterbewegung*).

Let us write Sommerfeld's formula:

$$(2.1) \qquad 1 + \frac{W}{m_0c^2} = \left[ 1 + \frac{Z^2\alpha^2}{\left[ n_r + (n_\varphi^2 - Z^2\alpha^2)^{1/2} \right]^2} \right]^{-1/2} \quad :$$

here: $W = E - m_0 c^2$; $n_r = 0,1,2,3,...$; $n_\varphi = 1,2,3,...$; we see that $n_\varphi = |\kappa|$: this difference between the azimuthal quantum number $n_\varphi$ and the auxiliary quantum number $\kappa$ must be ascribed to the spin influence: indeed, the values of $\kappa$ represent the eigenvalues of the following operator $D$ (see Dirac [1]), which is a constant of the motion:

$$(2.2) \qquad \hbar D := \rho_3 (\boldsymbol{\sigma} \cdot \mathbf{m} + \hbar),$$

where $\mathbf{m}$ and $(1/2)\hbar\boldsymbol{\sigma}$ are the orbital and the spin angular momentum respectively, and $\rho_3$ is a well-known operator of Clifford-Dirac algebra. (Our $D$ coincides with Dirac's *operator j* ).

Equation (2.1) tells us that when $Z\alpha \geq 1$ we encounter the *same* interpretative difficulty of equation (1.1): this is not a trivial remark, because it allows us to exclude any connection with the Klein paradox and the positron theory: we are confronted here with an essentially classical, i.e. non-quantal, difficulty. This can be seen in the clearest and explicit way with the following considerations.

First of all, we observe that it is easy to write the classical analogue of (2.1): it is sufficient to perform the following substitutions:

$$(2.3) \qquad \hbar n_\varphi \to p_\varphi \equiv p \; ; \quad \hbar n_r \to (1/2\pi) \oint p_r \, dr \equiv P \; ,$$

$$(2.3') \qquad p_\varphi = \frac{m_0 r^2 (d\varphi/dt)}{(1-\beta^2)^{1/2}} \; ; \quad p_r = \frac{m_0 (dr/dt)}{(1-\beta^2)^{1/2}} \; ; \quad (\beta^2 \equiv v^2/c^2) \; ;$$

$r$ and $\varphi$ are plane polar coordinates, $p$ and $P$ are constants of the motion and adiabatic invariants.

We obtain

$$(2.4) \qquad 1 + \frac{W}{m_0 c^2} = \left[ 1 + \frac{Z^2 e^4 / c^2}{[P + (p^2 - Z^2 e^4 / c^2)^{1/2}]^2} \right]^{-1/2} ;$$

clearly, eq. (2.4) makes sense only when $p^2 > Z^2 e^4 / c^2$, i.e. only when

$$(2.5) \qquad \gamma^2 := 1 - \frac{Z^2 e^4}{c^2} > 0 \; ;$$

(for $c \to \infty$, $\gamma^2 \to 1$, and eq. (2.4) gives the nonrelativistic energy). In other words, *for any value of Z formula* (2.4) *holds **only** for orbital angular momenta p greater than a minimal value* $p_{\min}(Z)$. An analysis of Sommerfeld's treatment [2] explains the reason of this restriction.

Sommerfeld started from the conservation theorems of the angular momentum $p_\varphi$ and of the energy $W$:

$$(2.6) \qquad p_\varphi \equiv p = \text{const.} \; ,$$

$$(2.7) \qquad \left[1 + \frac{1}{m_0 c^2}(p_r^2 + \frac{1}{r^2} p^2)\right]^{1/2} = 1 + \frac{W + Ze^2/r}{m_0 c^2} \; ;$$

now, if (and only if) $p$ is different from zero [i.e., if $(d\varphi/dt) \neq 0$], we can write

$$(2.8) \qquad \frac{p_r}{p} = \frac{(dr/dt)}{r^2 (d\varphi/dt)} = \frac{1}{r^2} \frac{dr}{d\varphi} \; ,$$

from which – with some manipulations – we get the differential equation of the electron trajectory $r = r(\varphi)$:

$$(2.9) \qquad \frac{d^2(1/r)}{d\varphi^2} + \gamma^2 \left(\frac{1}{r} - C\right) = 0 \; ,$$

$$(2.9') \qquad C := \frac{m_0 Z e^2}{p^2 \gamma^2} \left(1 + \frac{W}{m_0 c^2}\right) \; ;$$

the solution of (2.9) is, if $\varphi = 0$ is the angular coordinate of the perihelion $r = r_{\min}$:

$$(2.10) \qquad r = \frac{1/C}{1 + \varepsilon \cos \gamma \varphi} \; ,$$

where $\varepsilon$ is a constant of integration. In the nonrelativistic approximation ($\gamma^2 = 1$), eq. (2.10) represents a conic; if $W < 0$ we have an ellipse (*semilatus rectum* equal to $1/C$ and eccentricity $\varepsilon < 1$). For $0 < \gamma^2 < 1$ and $W < 0$, the orbit (2.10) is a "rosette" of precessing ellipses – a well-known pattern. (Remark that the rectilinear, swinging or non-swinging, trajectories for which $p = 0 = (d\varphi/dt)$ should be considered apart, both in the relativistic and nonrelativistic cases).

Then, Sommerfeld writes:

$$(2.11) \qquad p_r = \frac{\partial S}{\partial r} \;\; ; \;\; p_\varphi = \frac{\partial S}{\partial \varphi} \;\; ;$$

by substituting (2.11) into (2.7) we obtain the Hamilton-Jacobi equation of the relativistic Kepler problem:

$$(2.12) \qquad \left(\frac{\partial S}{\partial r}\right)^2 + \frac{1}{r^2}\left(\frac{\partial S}{\partial \varphi}\right)^2 = 2m_0 W + 2m_0 \frac{Ze^2}{r} + \frac{1}{c^2}\left(W + \frac{Ze^2}{r}\right)^2 \;\; ;$$

it is useful to remember that the Hamilton-Jacobi equation and the Dirac equation of the present problem are separable *only* in a polar frame.

We have $(\partial S/\partial \varphi) = p = \text{const.}$, and

$$(2.13) \qquad J_\varphi/(2\pi) := \frac{1}{2\pi}\int_0^{2\pi} p_\varphi d\varphi = \frac{1}{2\pi}\int_0^{2\pi} (\partial S/\partial \varphi)d\varphi = p \;\; ,$$

$$(2.14) \qquad J_r/(2\pi) := P := \frac{1}{2\pi}\oint p_r dr = \frac{1}{2\pi}\oint \frac{\partial S}{\partial r}dr \;\; .$$

Let us now substitute into (2.14) the expression of $\partial S/\partial r$ given by (2.12); the evaluation of the integral $\oint(\partial S/\partial r)dr$ yields the non-quantal formula (2.4). Finally, putting $p = n_\varphi \hbar$ and $P = n_r \hbar$, where $n_\varphi$ is a positive integer and $n_r$ a positive integer or zero, we arrive at Sommerfeld's formula (2.1).

**3.** − **Conclusion**

We have seen that the relativistic non-quantal formula (2.4) holds only for angular momenta $p$ greater than $Ze^2/c$. Quite similarly, the relativistic Sommerfeld's formula (2.1) holds only for azimuthal quantum numbers $n_\varphi$ greater than $Z\alpha$. Now, Dirac's formula (1.1) represents a refinement of Sommerfeld's result (2.1), i.e. a better description of physical reality. Accordingly, it is quite reasonable that Dirac's formula too is subjected to an analogous limitation: the absolute values of the auxiliary quantum number $\kappa$ must always be greater than $Z\alpha$.

The origin of all the above restrictions is *unique*: it is a "fault" of the special theory of relativity of *classical* physics.

‹‹ Es ist klar, daß ein Verständnis

der neuen Theorie nur auf der

Grundlage der älteren Theorie möglich ist ››.

A.Sommerfeld

**REFERENCES**

[1] See e.g.: WEYL H., *The Theory of Groups and Quantum Mechanics* (Dover Publications, New York, N.Y.) without indication of the year (likely 1931 or 1932), Ch.IV, sects. *7* and *8*; DIRAC P.A.M., *The Principles of Quantum Mechanics*, Fourth Edition (The Clarendon Press, Oxford) 1958, Ch. XI, sects. *71* and *72*.

[2] SOMMERFELD A., *Atombau und Spektrallinien – 1. Band,* Siebente Auflage (Friedr. Vieweg und Sohn, Braunschweig) 1951, Kap. *5*.

[3] FINKELNBURG W., *Structure of Matter* (Springer-Verlag, Berlin, *etc*.) 1964, pp.90 and 91.

[4] BJORKEN J.D. AND DRELL S.D., *Relativistic Quantum Mechanics* (McGraw-Hill Book Company, New York, *etc*.) 1964, p. 56.

[5] For a clear treatment of the Klein paradox see: PAULI W., *Helv. Phys. Acta,* **5** (1932) 179, sect.*4*; ID., *Handbuch der Physik, Band V/1* (Springer-Verlag, Berlin, *etc*.) 1958, Ziff.*24*.

[6] See sects. *1,2,3* of Pauli's paper [5], and Ziff.*23* of Pauli's article [5]. See also SOMMERFELD A., *Atombau und Spektrallinien – 2. Band,* Zweite Auflage (Friedr. Vieweg und Sohn, Braunschweig) 1951, p.714.